\documentclass[aps,pre,showpacs,floatfix,superscriptaddress,twocolumn]{revtex4}
\usepackage{graphicx}
\usepackage{amsmath}

\begin{document}

\title{Correlations in complex networks under attack}

\author{Animesh Srivastava}
\affiliation{Department of Computer Science and Engineering, Indian Institute of Technology Kharagpur, 721302 Kharagpur, India}

\author{Bivas Mitra} 
\affiliation{Universit{\'e} catholique de Louvain, Louvain-la-Neuve, Belgium}

\author{Niloy Ganguly}
\affiliation{Department of Computer Science and Engineering, Indian Institute of Technology Kharagpur, 721302 Kharagpur, India}

\author{Fernando Peruani} \email{peruani@unice.fr}
\affiliation{Laboratoire J.A. Dieudonn{\'e}, Universit{\'e} de Nice Sophia Antipolis, UMR CNRS 7351, Parc Valrose, F-06108 Nice Cedex 02, France}

\date{\today}

\begin{abstract}
For any initial correlated network after any kind of attack where either nodes or edges are removed, 
we obtain general  expressions for the degree-degree probability matrix and degree distribution. 
%
%
%
We show that  the proposed analytical approach predicts the correct topological
changes after the attack by comparing the evolution of the assortativity coefficient
for different attack strategies and intensities in theory and simulations.
We find that it is possible to turn an initial assortative network into a disassortative one, and vice versa, 
by fine-tuning removal of either nodes or edges. 
%
%
For an initial uncorrelated network, on the other hand, we discover that only a targeted edge-removal attack can
induce such correlations.
%
%
%
\end{abstract}


\maketitle

\section{Introduction}

The degree-degree correlations of a network is a critical property of the network topology.
For instance, these correlations, as well as the network degree distribution~\cite{new_gen,cohen_rand,calla}, play a crucial role on the resilience of the network~\cite{corr1, corr, corr2}.
Vazquez et al.~\cite{corr1} studied the impact of random node failure in
uncorrelated, assortative, and disassortative networks. They derived some general expressions to
show that the general criterion $\langle k^2 \rangle/\langle k \rangle \geq 2$ for percolation, derived explicitly for uncorrelated networks in~\cite{cohen_rand},  
is not applicable for  networks with degree-degree correlations.
%
In~\cite{corr}, Noh investigated numerically the nature of the percolation transition in correlated networks. 
His numerical results showed that
disassortative networks exhibit the same type of percolation transition as  neutral networks.
Recently, Goltsev et al.~\cite{corr2} contradicted Noh and demonstrated that both assortative and disassortative
mixing affect not only the percolation threshold but the critical behavior at the percolation transition point.
%
Their analysis showed that the critical behavior is determined by the eigenvalues of the
branching matrix and the degree-distribution.

The relevance of degree-degree correlations goes beyond just the network resilience.
These correlations also have a strong impact on the network dynamical properties as, for instance, its diffusion properties.
In correlated complex networks, the epidemic threshold  is determined by both, the degree distribution and degree-degree probability matrix~\cite{boguna2002, colizza2007}.
However, for some particular networks, as scale-free networks,  the epidemic threshold may not be affected by such correlations~\cite{boguna2003}. 
Despite this remarkable result, the (disease) spreading properties of most real world network are found to be extremely sensitive to degree-degree correlations~\cite{hufnagel2004, peruani2011}. 
Along similar lines, the removal of either node or edges can have a dramatic effect on the transport properties of a network as it has been shown to occur in the worldwide airport network~\cite{wu2006}. 
Certainly, if several airports are shut down, the circulation of passengers and goods will employ alternative routes and airports. 
The overload of edges and nodes may induce further damage in the network and eventually  a collapse of the entire transportation system. 
In \cite{wu2006} it was shown that the robustness of the worldwide airport network is particularly sensitive to node-node correlations. 
Here, we will learn that the removal of either nodes or edges affect the correlations themselves. 

The degree-degree correlations of a network can be characterized through an scalar:
the  assortativity (or Pearson) coefficient~\cite{aNewman2003,evolution_networks,reka_bara}.
This coefficient is zero when the network is uncorrelated. When it is
positive, it is said that the network is assortative. In assortative networks, most edges connect nodes that exhibit similar degree. 
On the other hand, disassortative networks, characterized by a negative
coefficien, are such that high-degree nodes are connected to low-degree
nodes.

Despite the relevance of degree-degree correlations, it has not been studied in detail how these correlations and their associated assortativity coefficient  are affected by an attack.
If we are able to predict the evolution of the degree distribution and degree-degree correlations after an attack, we will know most relevant feature of the distorted network such as
the new percolation threshold~\cite{new_gen,cohen_rand},   size of the giant component ~\cite{corr1, corr, corr2}, average path length~\cite{reka_bara},  or the new epidemic threshold~\cite{boguna2002, colizza2007}.

Here, we aim at filling this gap and focus on the effects that either a node- or an edge-removal attacks have on the degree distribution and degree-degree correlations of
a complex network.
More specifically,
we derive analytical expressions for the degree distribution and degree-degree
probability matrix of a correlated network under either node- or edge-removal attack.
We test the goodness of the analytical approach by simulating random and targeted attacks on initial networks which can be either assortative or disassortative (or neutral).
We compare the assortativity
coefficient obtained in theory and stochastic simulations
and find that the assortativity coefficient exhibits a non-trivial behavior with the attack intensity.
While random attacks, involving either node or edge removal, always reduce degree-degree correlations, targeted attacks
can induce drastic changes in the degree-degree probability matrix.
Interestingly, we find that such attacks can make an initial assortative network, disassortative, and vice versa.
For the particular case of an initial uncorrelated network, we find that only a targeted edge removal attack can induce correlations.

This paper is organized as follows. First we provide a more formal definition of the problem (Sec. \ref{sec:def}) then derive expressions
for the degree distribution and degree-degree probability matrix after a node removal attack, in Sec. \ref{sec:node}, and  after an edge removal attack,  in Sec. \ref{sec:edge}.
We present a comparison between stochastic simulations and the developed theory  in Sec. \ref{sec:comp}, and conclude in Sec. \ref{sec:conclu}.

\section{Problem definition} \label{sec:def}

Let us assume that the degree distribution of the initial network $p_i$ and
its degree-degree probability matrix $p_{i,j}$ are known.
Our goal is to obtain the degree distribution and the degree-degree
probability matrix after either node or edge removal attacks. We refer to these probabilities as $p'_i$ and
$p'_{i,j}$, respectively.
Notice that the degree-degree probability matrix contains the information
about the probability of finding an edge that connects a node of degree $i$
with another one of degree $j$, and obeys:
\begin{eqnarray}\label{eq:def_norm}
\sum_{i=0}^{k_{max}}\sum_{j=0}^{k_{max}} p_{i,j}  &=& 1\, ,\\
\label{eq:def_relation} \sum_{j=0}^{k_{max}} p_{i,j} &=& \frac{i p_i}{\langle k
    \rangle} \, ,
\end{eqnarray}
and for uncorrelated networks,
\begin{equation}
 p_{i,j} = \frac{ip_i}{\langle k\rangle}\frac{jp_j}{\langle k\rangle}\ ,
 \label{eq:pij_uncor}
\end{equation}
where $k_{max}$ denotes the maximum degree in the network, and $\langle k
    \rangle = \sum_{i} i p_i$. Eq. (\ref{eq:def_relation})
relates  $p_{i,j}$ and $p_{i}$. Similar expressions hold for $p'_i$ and
$p'_{i,j}$.

 \section{Impact of Node Removal Attacks} \label{sec:node}

We consider a generic node removal attack.
Let $f_k$ be the probability by which a node of degree $k$
is removed from the network.
Notice that $0 \leq f_k \leq 1$, and in general $\sum_k f_k \neq 1$.
This definition allows us to describe random and targeted (or deterministic) attacks.

Any node removal  attack can be thought as a process involving two steps.
The first step is to select the nodes that are going to be removed according
to the probability distribution $f_k$. After the selection of the nodes,
we divide the network into two subsets, one subset contains the nodes that are going to survive
 ($S$) while the other subset comprises of the nodes that are going
to be removed ($R$).  In the second step of the attack
all nodes in subset $R$ and all edges in $S$ that are linked to nodes in $R$
are removed.
We introduce the following definitions to facilitate further reading:
\begin{eqnarray}
n_{i,j}^{S,S} &=& p_{i,j} \langle k \rangle N (1-f_i) (1-f_j) \\
n_{i,j}^{S,R} &=& p_{i,j} \langle k \rangle N (1-f_i) f_j \\
n_{i,j}^{R,S} &=& p_{i,j} \langle k \rangle N f_i (1-f_j) \\
n_{i,j}^{R,R} &=& p_{i,j} \langle k \rangle N f_i f_j \, ,
\end{eqnarray}
where $n_{i,j}^{S,S}$ represents the number of tips -- let us recall that one edge has two tips -- that start from a node
of degree $i$ in $S$ and are connected to a tip which is linked to a node of degree $j$ also located in $S$, and
similarly for $n_{i,j}^{S,R}$, $n_{i,j}^{R,S}$ and $n_{i,j}^{R,R}$.


When the nodes in the subset $R$ are actually removed,
the degree distribution of the surviving nodes $S$ is changed due to the
removal of edges that run between the surviving set $S$ and any node of the
removed set $R$.
We focus on a node of degree $j$ in $S$ before the actual removal of nodes
in $R$.
We want to know the probability $\phi_j$ that one of the $j$ edges of this node is
connected to a node in $R$. This probability can be expressed as:
\begin{eqnarray} \label{eq:phij}
\phi_j = \frac{\sum_{k} n_{j,k}^{S,R}}{\sum_{k}
  (n_{j,k}^{S,R}+n_{j,k}^{S,S})} \, .
\end{eqnarray}
The removal of nodes can only lead to a decrease in the degree of a
survived node. If we find a node of degree $k$ that has survived, it
can be due to the fact that originally its degree was $k+q$ and $k$
of its edges survived, while $q$ ($q$ may be zero also) got removed. 
%
%
Hence, using Eq. (\ref{eq:phij}), we express $p_k'$ as the following binomial distribution:
\begin{equation}
 \label{eq:pk'}
 p'_k = \sum_{q=k}^\infty \left( \begin{array}{c} q \\ k
\end{array}\right) \phi_q^{q-k} (1-\phi_q)^{k}\, p^{s}_{q} \, ,
\end{equation}
where  $p_q^s = \frac{(1-f_q)p_q}{1-\sum_i p_if_i}$.

Notice that $\phi_j$ becomes independent of $j$ in two situations: a) when $f_k = f$,  and b) for
uncorrelated networks.
For $f_k = f$ (random node removal), $\phi_j = f$, while for uncorrelated networks
$\phi_j = \sum_k k\,p_k\,f_k/\langle k \rangle$.
In these two limiting cases, Eq.(\ref{eq:pk'}) reduces to the expression
derived in~\cite{bivaspre} for the degree distribution after the attack for
uncorrelated networks.
This means that the degree distribution after a random attack is independent of the
degree-degree correlations of the initial network and only depends on $p_k$.


Now, we look for a transformation that allows us to go from the initial degree-degree probability 
matrix to the joint degree probability matrix of the attacked network.
%
%
We know that the new matrix has to obey, by definition, Eq.(\ref{eq:def_relation}), i.e., $\sum_k
p'_{j,k} = j\,p'_j / \langle k' \rangle$
which implies a connection between the new degree-distribution and the new degree-degree probability matrix.
Taking this into account, let us focus on an edge that connects a node of degree $j$ and a node of degree $k$ in
the survived network.
Before the attack, these nodes have had a degree $\geq j$ and $\geq k$, respectively.  
This means that all edges that initially had an end connected to a node
of degree equal or larger than $j$, and the other end connected to a node of degree equal or larger than $k$, 
can contribute to the number of edges we observe after the attack connecting
nodes of degree $j$ and $k$.
Finally, if these two nodes are still
connected, then it is clear that the edge running between them before the
attack has also survived.
All this implies the following transformation:
%
%
%
\begin{eqnarray}
\label{eq:pij_p}
 p'_{j,k} = \sum_{u=j}^{k_{max}} \sum_{v=k}^{k_{max}} H(u,j, \phi_u) \cdot H(v,k, \phi_v)\cdot \xi_{u,v} \, ,
\end{eqnarray}
%
where to ease the notation we have introduced 
\begin{eqnarray} \label{eq:H} H(x,y, \omega)= \left(
\begin{array}{c} x-1 \\ y-1
\end{array}\right) \omega^{x-y} (1-\omega)^{y-1}\, ,
\end{eqnarray}
%
%
and defined $\xi_{u,v} = n_{u,v}^{S,S}/\sum_{m,l} n_{m,l}^{S,S}$, which is 
the probability of finding an edge connecting a node of degree $u$ and a node of degree $v$, both in the subset
$S$, before the attack.
It can be shown, through Eq.~(\ref{eq:pij_p}), that for an initial uncorrelated network that obeys  Eq. (\ref{eq:pij_uncor}), either a random or a targeted node removal attack leads to
 $p'_{j,k} = j\,p'_j\,k\,p'_k/\langle k' \rangle^2$ (Appendix~\ref{ap:NodeRem}).
Thus, a node removal attack can never correlate an initially uncorrelated network.
On the other hand, if the initial network exhibits correlations, a node removal attack will have an impact on the correlations~\footnote{As consequence of this, if a node removal attack is successively applied  on a network, and if at some point the distorted network becomes uncorrelated, it will remain uncorrelated. }.

%

\section{Impact of Edge Removal Attacks}\label{sec:edge}

In order to analyze the impact of link removal on the degree-distribution, we need first to establish a
relationship between the  degree-distribution $p_k'$ after the attack and the initial degree distribution
$p_k$ and degree-degree probability matrix $p_{i,j}$.
Let us represent by $f_{i,j}$ the probability that an edge, connecting nodes of degree $i$ and $j$, is removed during the attack.
The link removal attack is a two step process whereby first the edges to be removed are selected (with probability $f_{i,j}$) and then all the
selected edges are removed at once. It is important to note that unlike node removal attacks, link removal attack
does not divide the network into two subsets.


For an undirected network, an edge between any two nodes $u$ and $v$ can be thought of as a set of two edges:
from $u$ to $v$ and from $v$ to $u$.
Hence, the total number of edges in this ``undirected'' network
is given by $N\langle k\rangle$, where $N$ is the number of nodes in the network,  $\langle k\rangle$ is the mean degree, and the total number of edges
from $i$-degree nodes and to $j$-degree nodes is given by $N\langle k\rangle p_{i,j}$.
Out of these many edges, $N\langle k\rangle p_{i,j}f_{i,j}$ edges will be removed.
This
helps us to derive the total number of removed edges whose one end is connected to an $i$-degree node, while the other end is connected
to any other degree node, which can be expressed as
\begin{equation}
 \label{eq:Ei}
 E_i = N\langle k\rangle\sum_j p_{i,j}f_{i,j} \, .
\end{equation}
The quantity $E_i$ represents the number of tips which connect to $i$-degree nodes that are removed. This quantity can be used to compute
$\tilde{\phi}_i$, the probability that a node of degree $i$ loses a tip, which reads:
\begin{equation}
 \label{eq:Phii}
 \tilde{\phi}_i = \frac{E_i}{iNp_i} \, .
\end{equation}
The removal of edges can only lead to a decrease in the degree of a node. If we find a node of degree $k$ after the attack,
it can be due to the fact that originally its degree was $q$, with $k\leq q\leq k_{max}$, and $k$ of its edges survived, while $q-k$
got removed.
Thus, from the Eqs.~(\ref{eq:Ei}) and~(\ref{eq:Phii}), and assuming that the edges of a node are independent, we
obtain the following expression for $p_k'$:
\begin{equation}
 \label{eq:pkdash}
 p_k' = \sum_{q=k}^{k_{max}} \left( \begin{array}{c} q \\ k \end{array}\right) \tilde{\phi}_q^{q-k} (1-\tilde{\phi}_q)^{k}\, p_q\ \, .
\end{equation}
Notice that  for $f_{i,j} = f$, $\tilde{\phi}_q$ becomes independent of $q$.
On the other hand, for uncorrelated networks,
$\tilde{\phi}_q$ reduces to $\sum_k k\,p_{k}\,f_{q,k}/\langle k\rangle$.


In the following we derive an expression for the degree-degree probability matrix $p'_{j,k}$ after the attack.
Given an edge removal attack characterized by $f_{i,j}$, we look for a transformation that allows us to move from the initial degree-degree probability matrix
$p_{j,k}$ to the new probability matrix $p'_{j,k}$, which has to obey Eqs.~(\ref{eq:def_norm})
and~(\ref{eq:def_relation}).
If we find an edge connecting nodes of degree $j$ and $k$ in the network after the attack,
we can assume that before the attack the edge was connecting nodes of degree $u$ and $v$,
with  $j \leq u \leq k_{max}$ and $k \leq v \leq k_{max}$.
Since the selected edge is not removed from the network, then 
this means that the initial $u$-degree node lost $u-j$ edges (from its initial $u-1$ edges not linked to the analyzed edge), while the $v$ degree node lost $v-k$ edges. 
As result of this process,  the node degree after the attack is $j$ and $k$, respectively. 
In consequence,  using the probability 
$\tilde{\phi}_k$ given by Eq.~(\ref{eq:Phii}), we can express the degree-degree probability matrix after the edge removal attack as:
\begin{eqnarray}
 \label{eq:corr_mat_dash}
 p'_{j,k} = \sum_{u=j}^{k_{max}} \sum_{v=k}^{k_{max}} H(u,j,\tilde{\phi}_u) \cdot H(v,k,\tilde{\phi}_v)\cdot p_{u,v},
\end{eqnarray}
where $H(x,y,\omega)$ is again given by Eq.~(\ref{eq:H}). 
It can be shown that for an initial uncorrelated network that obeys  Eq. (\ref{eq:pij_uncor}), a random edge removal attack leads, according to Eq.~(\ref{eq:corr_mat_dash}),  to
 $p'_{j,k} = j\,p'_j\,k\,p'_k/\langle k' \rangle^2$ (Appendix \ref{ap:EdgeRem}).
This means that the random removal of edges cannot correlate an initial uncorrelated network.
On the contrary, a targeted edge removal attack can induce correlations in an initial uncorrelated network. The proof is given in Appendix~\ref{ap:EdgeRem}. 
If, on the other hand, the initial network is correlated, both a random or a targeted edge removal attack will affect the network correlations.

\section{Comparison between theory and stochastic simulations}\label{sec:comp}

We test the goodness of the analytical approach by comparing the degree distribution and the degree-degree
probability matrix obtained from the theory and from stochastic simulations.
The comparison is performed through the assortativity coefficient $r$ that is defined as follows~\cite{assort}: 
\begin{eqnarray} \label{eq:assortiativity_coeff}
r =&& \frac{\sum_{j,k} j\,k\,p_{j,k} - \left( \sum_{j,k} \frac{(j+k)}{2}\,
    p_{j,k}  \right)^2}{\sum_{j,k} \frac{(j^2 + k^2)}{2} \, p_{j,k} - \left( \sum_{j,k} \frac{(j+k)}{2}\,
    p_{j,k}  \right)^2 } \, .
\end{eqnarray}
The following convention is used:  `` $r$ '' refers to the  initial assortativity coefficient, while  ``$r'$" to the coefficient after the attack. 
Thus, $r'$ is a function of $p'_{j,k}$, see Eqs. (\ref{eq:pij_p}) and (\ref{eq:corr_mat_dash}).

The comparison has been performed on Erdos-Renyi, bimodal, and scale-free networks, 
obtaining in all cases an excellent agreement between theory and simulations. 
To illustrate the goodness of theory, we choose to present only results on scale-free networks given their broad applicability. 
The various attacks were simulated on initial assortative and disassortative scale-free networks. 
Thus, we refer  to these initial networks as $IDN$ (Initial Disassortative Network) and $IAN$ (Initial Assortative Network). 
These networks were generated using the method described in ~\cite{tabourier2011generating} and their details are given below.

\begin{enumerate}
\item  $IDN$ is characterized by a negative assortativity coefficient $r=-0.168$, and a power-law degree distribution of exponent  $-2.3$.
The first and second moments of the degree-distribution are
$\langle k \rangle = 3.2348$ and $\langle k^2 \rangle = 28.9350$, respectively, and maximum degree $37$. 
\item $IAN$ is characterized by a positive assortativity coefficient $r=0.275$, and a power-law degree distribution of exponent  $-2.3$.
The first and second moments of the degree-distribution are
$\langle k \rangle = 2.3760$ and $\langle k^2 \rangle = 11.2140$, respectively, and maximum degree $23$. 
\end{enumerate}

\subsection{Results for node removal}
\begin{figure}[t]
\begin{center}
\includegraphics[width=\columnwidth]{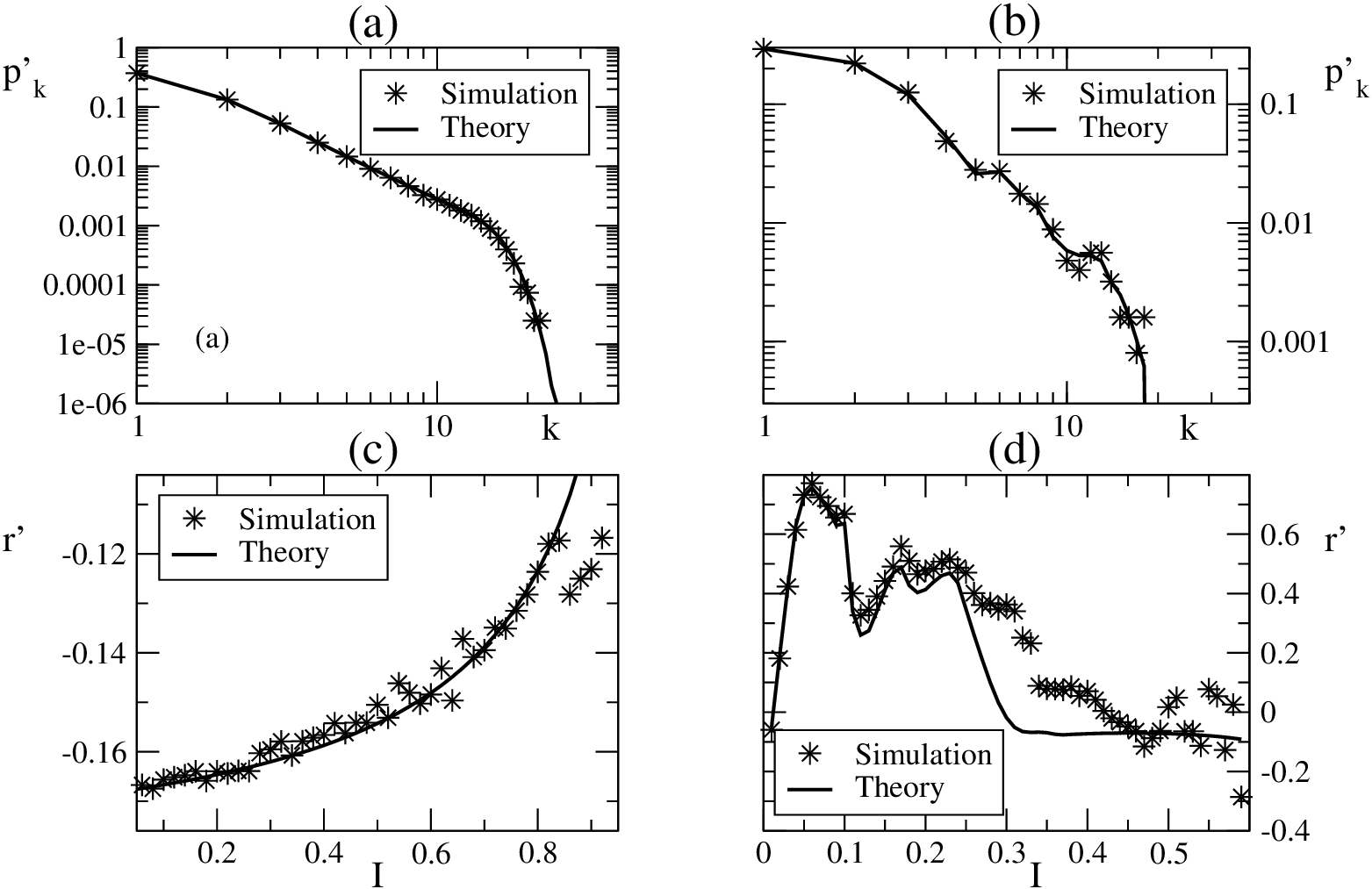}
\caption{Impact of {\it node} removal attacks on $IDN$.  (a) and (b) show the degree distribution of the attacked network after the random removal of $60\%$ of its
nodes, in (a),  and the removal of the $2\%$ of the highest degree nodes, in (b). The solid lines correspond to Eq.~(\ref{eq:pk'}).
(c) and (d) show the  assortativity coefficient $r'$ as function of the attack intensity $I$ for random  and targeted attack, respectively. The solid curves correspond to the evaluation of
 Eq.~(\ref{eq:assortiativity_coeff}) using Eq.~(\ref{eq:pij_p}).
}
\label{fig:NetANodeAll}
\end{center}
\end{figure}
We tested two node removal attacks~\cite{holme2002}: a random attack, sometimes also referred to as failure, where $f_k = f$,
and a targeted attack given by:
\begin{equation} \label{eq:detAttack}
f_k =
\begin{cases}
1 & \mbox{for } k> k_{cut}  \\
q & \mbox{for } k= k_{cut} \\
0 & \mbox{for } k< k_{cut} \\
\end{cases} \, .
\end{equation}
The first attack defines a situation in which randomly selected nodes are removed from the network,
independent of their degree.
The second attack defines a targeted attack procedure where all nodes
having  degrees higher than $k_{cut}$ are removed.
The attack intensity $I$ of a node-removal attack is given by the fraction of nodes that are removed from network.
For $f_k=f$, $I=f$.
For the attack given by Eq.~(\ref{eq:detAttack}), $I = q\cdot p_{k_{cut}} + \sum_{k=k_{cut}+1}^{k_{max}} p_k$.

\begin{figure}[t]
\begin{center}
\includegraphics[width=\columnwidth]{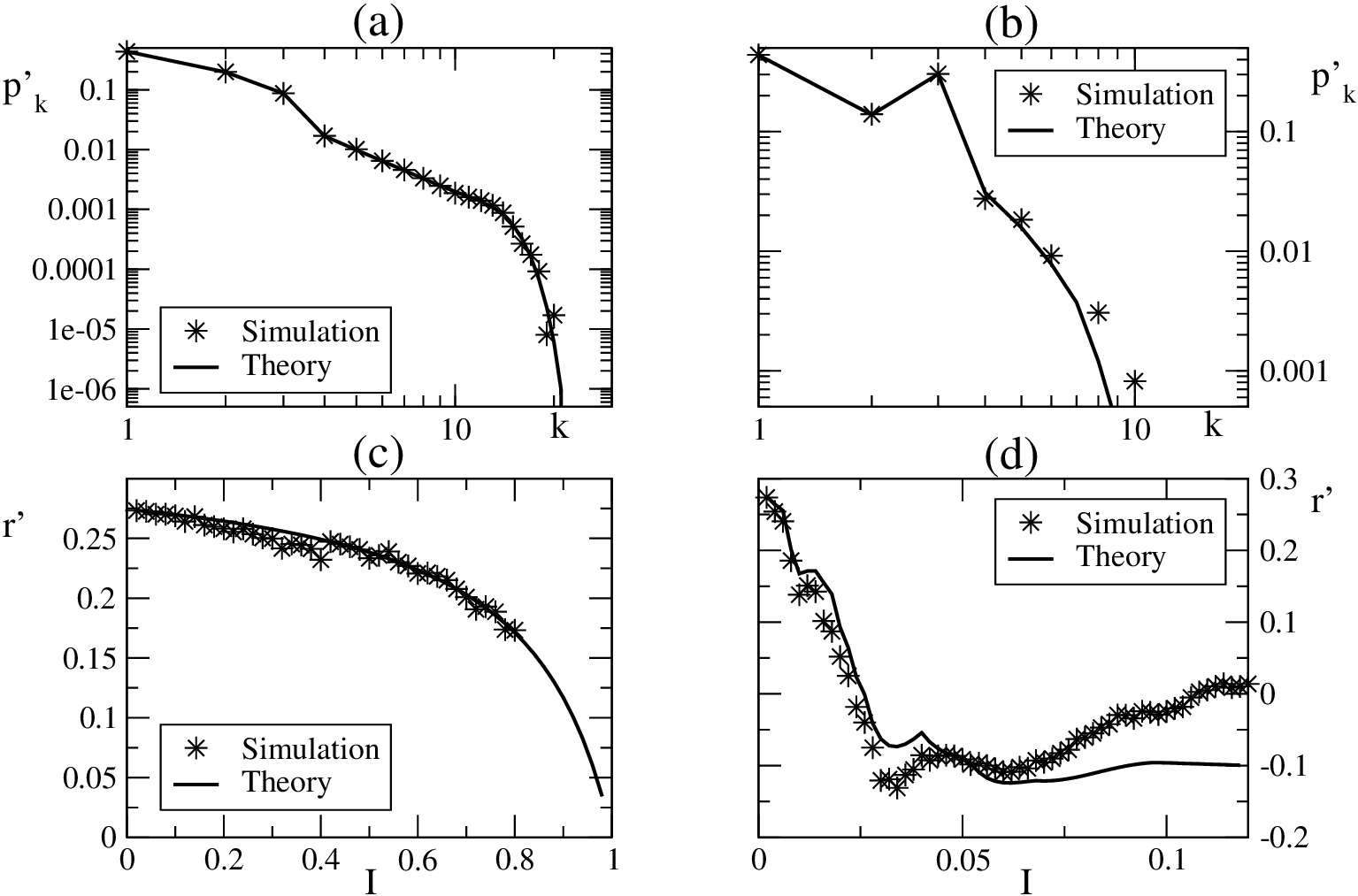}
\caption{
Impact of {\it node} removal attacks on $IAN$.  (a) and (b) show the degree distribution of the attacked network after the random removal of $40\%$ of its
nodes, in (a),  and the removal of the $2\%$ of the highest degree nodes, in (b). The solid lines correspond to Eq.~(\ref{eq:pk'}).
(c) and (d) show the  assortativity coefficient $r'$ as function of the attack intensity $I$ for random  and targeted attack, respectively. The solid curves correspond to the evaluation of
 Eq.~(\ref{eq:assortiativity_coeff}) using Eq.~(\ref{eq:pij_p}).
}
\label{fig:NetBNodeAll}
\end{center}
\end{figure}

Figs.~\ref{fig:NetANodeAll}~$(a)$ and $(b)$  and Fig.~\ref{fig:NetBNodeAll} ~$(a)$ and $(b)$  show that
 for an initial either disassortative or assortative network, Eq.~(\ref{eq:pk'})   predicts the correct deformed degree distribution. 
 On the other hand, Figs.~\ref{fig:NetANodeAll}~$(c)$ and \ref{fig:NetBNodeAll} ~$(c)$ indicate that for a random attack Eq. (\ref{eq:pij_p}) allows us 
 to compute the correct assortativity coefficient for both, disassortative and assortative initial networks. 
 From the figures it can be inferred  that random removal of nodes induces randomness, and consequently the assortativity 
 coefficient $r'$ of $IDN$ increases as the attack intensity is increased, while for $IAN$, $r'$ decreases. 
 In both cases, $r' \to 0$ as $I \to 1$.  
 In case of a targeted attack, the network correlations exhibit a complex, non trivial behavior with ups and downs as the attack intensity is increased, 
see Figs.~\ref{fig:NetANodeAll}~$(d)$ and \ref{fig:NetBNodeAll}~$(d)$. 
This complex behavior is also predicted by Eq. (\ref{eq:pij_p}) which indicates that the observed complex functional form of $r'$ with $I$ is not due to 
arbitrary fluctuations that result from a poor statistics. 
Below we offer a tentative, more physical interpretation of these non trivial curves.

A targeted node-removal attack affects firstly high degree nodes.  
For an initial disassortative network ($IDN$) the removal of few of the highest degree nodes leads to an homogenization of the network.  
Nodes tend to have similar degree, and so most connection occur among nodes that exhibit similar degree. 
Consequently, the assortative coefficient increases, becoming even positive; see maximum in Fig.~\ref{fig:NetANodeAll}$(d)$.
Further removal of nodes has the opposite effect. This is arguably due to the fact that after most hubs in the system have been removed, 
a targeted attack is not very different than a random attack, and so $r'$ decreases.  
This latter observation can be easily understood if we imagine a simpler scenario where all nodes have the same degree. 
Then, the removal of nodes necessarily induces heterogeneity and  $r'$ tends to $0$. 
Interestingly, $r'(I)$ in  Fig.~\ref{fig:NetANodeAll}$(d)$ is more complex than what we have just described. 
Particularly intriguing is the fact that for large $I$ values $r'$ seems to tend asymptotically to a small but negative value. 
Though the arguments provided above do not account for all the details of the curve, they constitute a tentative explanation for the 
non monotonic shape of the curve and the observed transition from disassortative to assortative. 

The non-monotonic behavior observed in Fig.~\ref{fig:NetBNodeAll}$(d)$ for an initial assortative network, $IAN$, can be understood along similar lines. 
The removal of few high-degree nodes, leads to a dramatic reduction of the number of edges running among high-degree nodes. 
Consequently, the statistical weight of those connections running between high and low-degree nodes can become remarkably  important.  
Fig.~\ref{fig:NetBNodeAll}$(d)$ clearly shows that targeted node removal can even make the assortative coefficient for an IAN become negative.

Fig.~\ref{fig:NetANodeAll} and ~\ref{fig:NetBNodeAll} show that
despite of the complexity of the process,  Eq.~(\ref{eq:pij_p}) is able to predict the correct degree-degree correlations of the network after the attack.
Deviations between the theory and simulations are observed  when some simulation attacks started to lead to heavily fragmented networks.
As the number of simulation attacks is increased, the agreement between Eq.~(\ref{eq:pij_p})  and the numerically obtained $r'$  seems to become systematically better. 
More importantly, we have learned that  a targeted node removal attack can be used to transform an initial assortative network into a disassortative, and vice versa (see Figs. \ref{fig:NetANodeAll}(d) and \ref{fig:NetBNodeAll}(d)).

\begin{figure}[t]
 \begin{center}
  \includegraphics[width=\columnwidth]{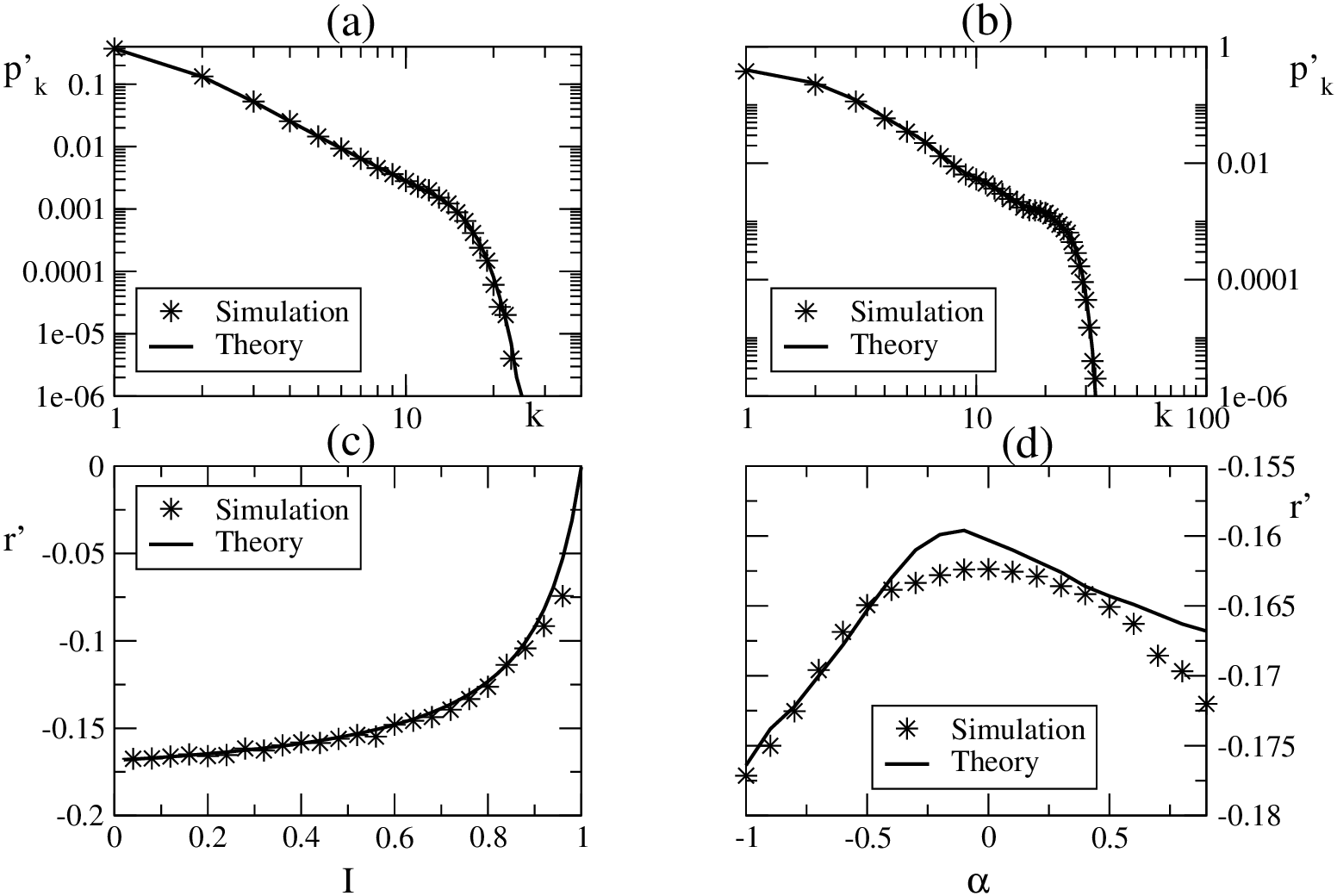}
 \end{center}
\caption{
Impact of {\it edge} removal attacks on $IDN$.  (a) and (b) show the degree distribution of the attacked network after the random removal of $60\%$ of its
edges, in (a),  and the targeted removal of $20\%$ edges given by Eq.~(\ref{eq:detEdgeAttack}) using $\alpha=0.5$, in (b).
The solid lines correspond to Eq.~(\ref{eq:pkdash}).
(c) shows the  assortativity coefficient $r'$ as function of the random edge attack intensity $I$, while (d) corresponds to $r'$, due to the
removal of $3\%$ of edges, as function of the targeted attack parameter $\alpha$.
The solid curves correspond to the evaluation of  Eq.~(\ref{eq:assortiativity_coeff}) using Eq.~(\ref{eq:corr_mat_dash}).
}
\label{fig:NetAAllEdge}
\end{figure}

\begin{figure}[t]
 \begin{center}
  \includegraphics[width=\columnwidth]{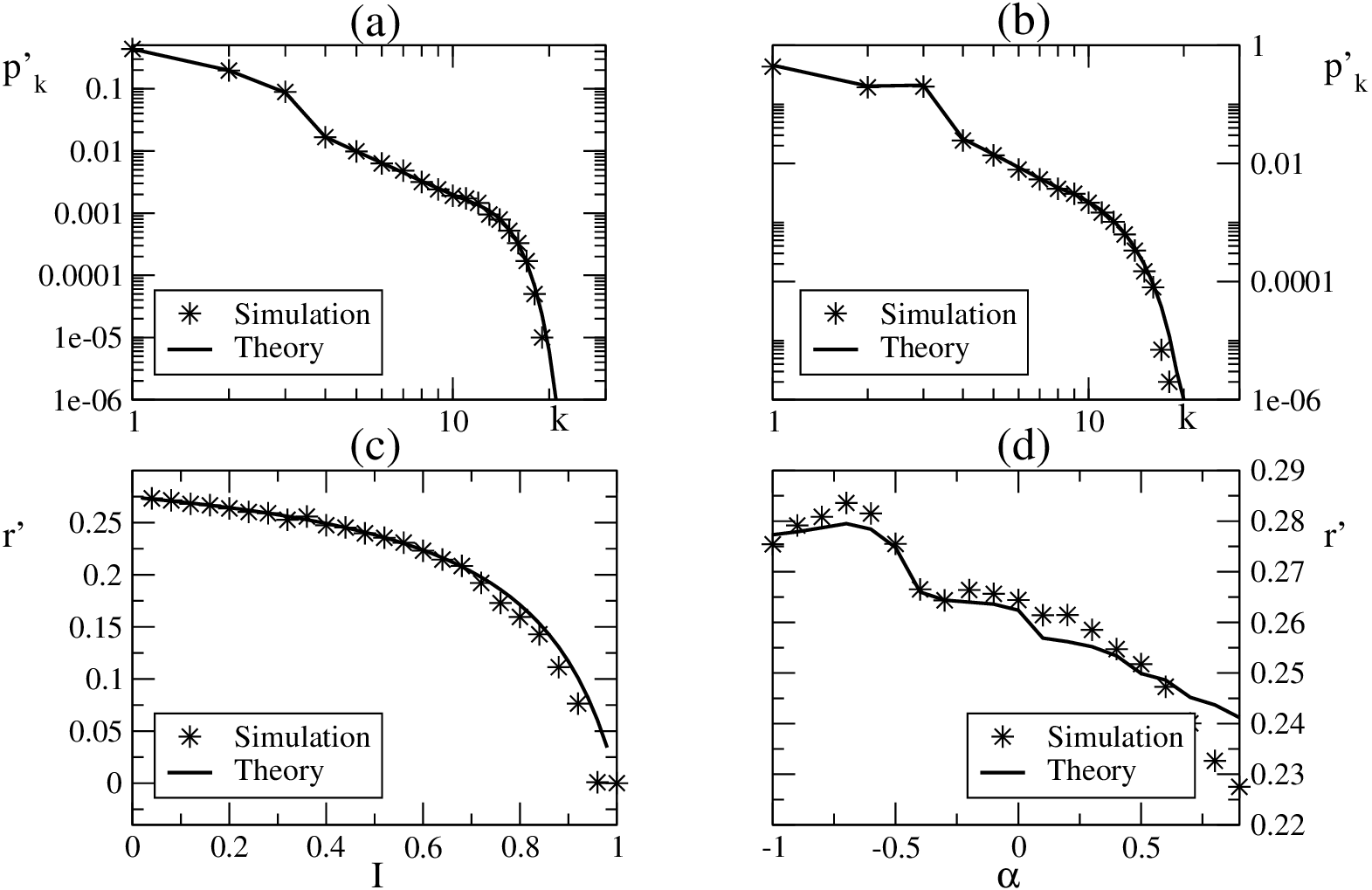}
 \end{center}
\caption{
Impact of {\it edge} removal attacks on $IAN$.  (a) and (b) show the degree distribution of the attacked network after the random removal of $40\%$ of its
edges, in (a),  and the targeted removal of $20\%$ edges given by Eq.~(\ref{eq:detEdgeAttack}) using $\alpha=0.5$, in (b).
The solid lines correspond to Eq.~(\ref{eq:pkdash}).
(c) shows the  assortativity coefficient $r'$ as function of the random edge attack intensity $I$, while (d) corresponds to $r'$,
due to the removal of $5\%$ edges, as function of the targeted attack parameter $\alpha$.
The solid curves correspond to the evaluation of  Eq.~(\ref{eq:assortiativity_coeff}) using Eq.~(\ref{eq:corr_mat_dash}).
}
\label{fig:NetBAllEdge}
\end{figure}

\subsection{Results for edge removal}

We tested two attacks for edge removal:  a random attack,  with $f_{i,j} = f$, and a targeted attack~\cite{moreira2009} of the form:
\begin{equation} \label{eq:detEdgeAttack}
f_{i,j} = \beta\,(i \, j)^{\alpha} \, ,
\end{equation}
where $\beta$ is a normalization constant and $\alpha$ is another constant that controls the type of attack (see also~\cite{wu2006}). 
%
%
Notice that for the same number of removed edges, different values of  $\alpha$ induce  different effects.
For instance,  with $\alpha=1$, the attack tends to affect those  edges connecting high degree nodes, while for
$\alpha=-1$ edges running between low degree nodes are more likely to be removed.
For targeted edge removal attack, we have conducted attack simulations using as control parameter  the (attack) exponent $\alpha$, while 
keeping constant the number of edges to be removed, i.e. $I=const$.  
For random edge-removal attacks, the attack intensity $I$ is simply  $I=f$, while for a targeted edge-removal attack, $I=\sum_{i,j} p_{i,j} f_{i,j}$.

Figs.~\ref{fig:NetAAllEdge}~$(a)$ and $(b)$, and Figs.~\ref{fig:NetBAllEdge}~$(a)$ and $(b)$ show that Eq.~(\ref{eq:pkdash})
suffices to predict the new degree distribution after edge removal attacks for either a  disassortative  (Fig.~\ref{fig:NetAAllEdge}) or assortative (Fig.~\ref{fig:NetBAllEdge}) initial network.
From Fig.~\ref{fig:NetAAllEdge}~$(c)$ and Fig.~\ref{fig:NetBAllEdge}~$(c)$, we learn that random edge removal, $f_{i,j}=f$,
 induces randomness, with the assortativity coefficient getting closer to  $0$ as the attack intensifies.
%
%
In summary, a random attack, involving either node or edge removal, always weakens the degree-degree correlations exhibited by the initial network.

A targeted attack, on the other hand, can introduce new correlations in the  network.
Figs.~\ref{fig:NetAAllEdge}~$(d)$ and~\ref{fig:NetBAllEdge}~$(d)$ show the impact of targeted edge removal on
the degree-degree probability matrix of $IDN$ and $IAN$ with respect to the attack parameter $\alpha$ for a given value of $I$.  
%
%
%
For an initial disassortative network, $IDN$, as the attack
parameter $\alpha$ changes from $0$ to $-1$, the removal of edges running between low-degree nodes gets intensified. 
The removal of these edges  leads to the removal of low-degree nodes from the network. This implies   
an increase in the proportion of edges connecting low-degree and high-degree nodes in the deformed network. 
At the same time, the attack reduces  the
degree of a low-degree nodes, which  induces a further decrease of $r'$
%
%
This dynamics  explains the decrease of $r'$ observed in Figs.~\ref{fig:NetAAllEdge}~$(d)$ as we move from $\alpha=0$ to $-1$.  
For $0 \leq \alpha \leq 1$,  the attack affects those edges running between  high-degree nodes. Hence,  the fraction of
edges connecting high-degree  and low-degree nodes increases as $\alpha \to 1$ and $r'$ gets more negative, as observed in Figs.~\ref{fig:NetAAllEdge}~$(d)$.  
In summary, for a disassortative initial network ($IDN$) any targeted removal of edges at constant $I$ (either with $\alpha>0$ or $\alpha<0$) seems to increase negative correlations, making the network more disassortative. 


For an initial assortative network, $IAN$, we observe that as $\alpha$ is increased from $0$ to $1$, $r'$ decreases, Figs.~\ref{fig:NetBAllEdge}~$(d)$. 
Arguably, this is due to the removal of some of the abundant  edges connecting high-degree nodes, which brings in randomness. 
On the other hand,  as $\alpha$ is decreased from $0$ to $-1$, several edges connecting low degree nodes are removed, which  leads to the removal of 
low degree nodes. 
As the network becomes more homogeneous, $r'$ exhibits an increase,  Figs.~\ref{fig:NetBAllEdge}~$(d)$. 
Though not shown here, by increasing the attack intensity, a targeted edge removal attack can be used to transform an IAN into a disassortative one.
While the opposite could be  possible, we have not observed it numerically.  
Despite the apparent complexity of process, Figs.~\ref{fig:NetAAllEdge}~$(c)$ and $(d)$,
and Figs.~\ref{fig:NetBAllEdge}~$(c)$ and $(d)$ show that Eq.~(\ref{eq:corr_mat_dash}) is able to predict the correlation
changes due to the different attacks for either disassortiative or assortative initial networks.

Table~\ref{table:summ}
summarizes the results obtained for both, edge and node removal attacks. 
Numerical and analytical results for the particular case of an  initial uncorrelated network are given in the appendix.

\begin{center}
\begin{table}
\begin{tabular}{c | c | c | c | c}
 \hline
         &  \multicolumn{2}{c |}{Node-Removals}  & \multicolumn{2}{c}{Edge-Removals} \\
 \cline{2-5}
 Initial Network & Random & Targeted & Random & Targeted\\
 \cline{1-5}
 IDN     & RC & IC & RC  & IC \\
 IAN     & RC & IC & RC & IC \\
 IUN  & N & N & N & IC \\
 \cline{1-5}
\end{tabular}
\caption{The table summarize the obtained results. Starting from an initial network which can be disassortative (IDN), assortative (IAN), 
or uncorrelated (IUN), we indicate whether the corresponding attack induces correlations (IC), removes correlations (RC),  or whether it does not affect the network correlations (N). 
}
\label{table:summ}
\end{table}
\end{center}

\section{Conclusions}\label{sec:conclu}
%

We have derived an analytical framework that has allowed us to  understand the impact of node- and edge-removal
attacks on the correlations of complex networks. Stochastic simulation results indicate
that the derived theory provides a good estimate of the degree distribution and degree-degree probability matrix under
node- and edge-removal attacks for both, assortative and disassortative initial networks. The main insights
obtained from this work are: 
\begin{enumerate}
\item Random node- or edge-removals always introduce randomness in the deformed network which tends to become uncorrelated as the attack intensity is increased. 
\item Targeted node- or edge-removals can strongly affect the network correlations to the point that an initial assortative network can turn into disassortative, and vice versa.
\item If the initial network is
uncorrelated, only a targeted edge removal attack can introduce correlations. All other attacks defined in this paper
keep the network uncorrelated.
\end{enumerate}

These results, beyond their academic interest, are relevant from a practical point of view. 
As briefly explained in the introduction, degree-degree correlations control several network properties as robustness~\cite{corr1, corr, corr2}, path length~\cite{reka_bara}, 
and diffusive properties~\cite{boguna2002, colizza2007, wu2006, peruani2011}, among many others. 
As we have shown here, targeted as well as unintentional removal of either nodes or edges affect the degree-degree correlations. 
Consequently, the above mentioned properties -- robustness, path length, etc -- are also affected. 
Since most of these properties are known functions of $p'_i$ and $p'_{i,j}$, the expressions derived here -- Eqs. (\ref{eq:pk'}),  (\ref{eq:pij_p}), (\ref{eq:pkdash}), and (\ref{eq:corr_mat_dash}) --  are useful tools that allow us to recompute all these quantities for a correlated network subject to any type of attack. 
%

%

\section*{Acknowledgments}

We thank L. Tabourier for useful discussions on generating constrained random graphs. F.P. acknowledges the
hospitality of IIT-Kharagpur. This work has been  partially supported by STIC-Asie project titled ``Information
spreading in a system of mobile agents", from the French Government. This work has also been partially funded by
a project titled ``Building Delay Tolerant Peer to Peer Network'' under Department of Information and Technology
(DIT), Govt. of India.


\appendix

\section{Node-removal attack on an initial uncorrelated network}\label{ap:NodeRem}

\begin{figure}
 \begin{center}
  \includegraphics[width=\columnwidth]{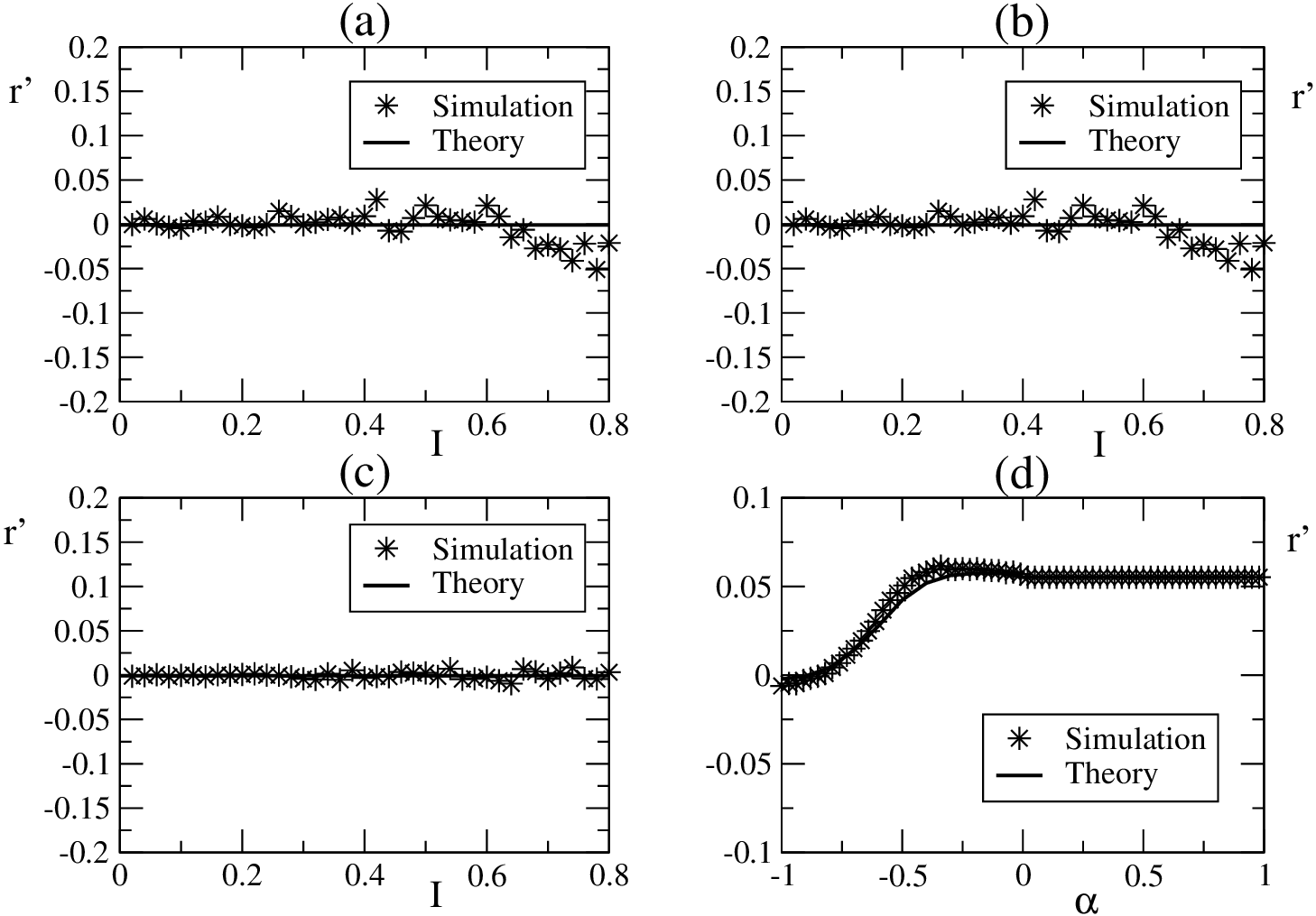}
 \end{center}
 \caption{Change in assortativity of an initial uncorrelated network due to (a) random node-removal, (b) targeted node-removal,
 (c) random edge-removal, and (d) targeted edge-removal attack. In (d), the attack intensity corresponds to the removal  of $3\%$ of edges. Notice that only a targeted edge-removal attack is able to affect the assortativity coefficient.}
 \label{fig:UncorrNodeEdge}
\end{figure}

Here, we show that for an initial uncorrelated network, a node-removal attack cannot induce correlations.
The probability $p_{i,j}$ of the initial network is then given by Eq.~(\ref{eq:pij_uncor}).
In consequence, the probability that a node loses a tip, given by Eq.~(\ref{eq:phij}), becomes independent of its degree:
$\phi_j=\phi$. Eq.~(\ref{eq:pij_uncor}) also implies that the probability $\xi_{u,v}$ of finding an edge between nodes of degree $u$ and $v$ in set $S$ becomes
\begin{equation}
 \label{eq:xirand}
 \xi_{u,v} = \frac{u\,p_u\,v\,p_v\,(1-f_u)\,(1-f_v)}{\sum_u\sum_v u\,p_u\,v\,p_v\,(1-f_u)\,(1-f_v)} \,.
\end{equation}
The average degree of the deformed network, $\langle k'\rangle = \sum_k k\,{p'}_{k}$, and can then be expressed as
\begin{eqnarray}
 \label{eq:defk}
 \langle k'\rangle
           &=& \sum_{k=0}^{k_{max}} k\sum_{q=k}^{k_{max}} \left( \begin{array}{c} q \\ k\end{array}\right) \phi^{q-k} (1-\phi)^{k}\, \frac{p_{q}(1-f_q)}{\sum_k p_k(1-f_k)} \nonumber\\
           &=& \sum_{q=0}^{k_{max}} \frac{q\,p_q\,(1-f_q)\,(1-\phi)}{\sum_k p_k(1-f_k)}\cdot\nonumber\\
                                            && \sum_{k=0}^{q} ( \begin{array}{c} q-1\\k-1\end{array})\phi^{q-k}(1-\phi)^{k-1} \nonumber\\
           &=&  \frac{\sum_{q=0}^{k_{max}} q\,p_q\,(1-f_q)\,(1-\phi)}{\sum_k p_k(1-f_k)}
\end{eqnarray}
Using $\phi_u = \phi_v = \phi$ and Eqs.~(\ref{eq:xirand}) and~(\ref{eq:defk}), Eq.~(\ref{eq:pij_p}) reduces to
\begin{eqnarray}
 \label{eq:pij_p_rand}
 p'_{i,j} &=& \frac{i\,p'_i\,j\,p'_j}{\left[\frac{(1-\phi)\sum_k k\,p_k\,(1-f_k)}{\sum_k p_k\,(1-f_k)}\right]^2} \nonumber\\
          &=& \frac{i\,p'_i\,j\,p'_j}{\langle k'\rangle^2} \, .
\end{eqnarray}
This implies that under any kind of node-based attack, an initially random network remains random. This has also been observed in
simulations as shown in Figs.~\ref{fig:UncorrNodeEdge}~$(a)$ and $(b)$.

\section{Edge-removal attack on an initial uncorrelated network}\label{ap:EdgeRem}
Here, we show that for an initial uncorrelated network, a random edge-removal attack cannot induce correlations, while, on the contrary, a targeted edge-removal attack can do it.
For an initial uncorrelated network, $p_{i,j}$ is given by Eq.~(\ref{eq:pij_uncor}), and then the probability that a node of degree $i$ loses a tip, given by
 Eq.~(\ref{eq:Phii}), reduces to:
\begin{equation}
 \label{eq:phiedgerand}
 \tilde{\phi}_i = \frac{\sum_j j\,p_j\,f_{i,j}}{\langle k\rangle} \, .
\end{equation}
Using Eq.~(\ref{eq:pij_uncor}) in Eq.~(\ref{eq:corr_mat_dash}), the probability that an edge exists between nodes of degree $i$
and $j$ in the deformed network can be expressed as
\begin{eqnarray}
 \label{eq:corr_mat_dash_rand}
 p'_{i,j} = \frac{i\, j}{\langle k\rangle^2}\sum_{u=i}^{k_{max}}\frac{u}{i}\,H(u,i,\tilde{\phi}_u)\,p_u\sum_{v=j}^{k_{max}}\frac{v}{j}\,H(v,j,\tilde{\phi}_v)\,p_v \, ,
\end{eqnarray}
where $H$ is given by Eq. (\ref{eq:H}).  
In the case of random edge attack, $f_{i,j}=f$,  and so Eq.~(\ref{eq:phiedgerand})  becomes
$\tilde{\phi}_u=f$. Using this in Eq.~(\ref{eq:corr_mat_dash_rand}) we get
\begin{equation}
 \label{eq:corr_mat_dash_rand2}
 p'_{i,j} = \frac{i\,p'_i\,j\,p'_j}{[\sum_k k\,p_k\,(1-f)]^2} = \frac{i\,p'_i\,j\,p'_j}{\langle k'\rangle^2}\, .
\end{equation}

This shows that under random edge removal attack an initially uncorrelated network remains uncorrelated, see Fig. ~\ref{fig:UncorrNodeEdge}(c). 
In case of targeted edge removal,  $\tilde{\phi}_i$ does not become  degree independent and
Eq.~(\ref{eq:corr_mat_dash_rand}) does not get reduced to Eq.~(\ref{eq:pij_uncor}). 
This implies that correlations  have crept in the attacked network as Fig. \ref{fig:UncorrNodeEdge}~$(d)$ confirms. 
Therefore, it is only through a targeted edge removal attack that degree-degree correlations can be induced in an initial uncorrelated network.

%


\bibliographystyle{apsrev}



\end{document}